\documentclass{article}

\usepackage{arxiv}

\usepackage[utf8]{inputenc} 
\usepackage[T1]{fontenc}    
\usepackage{hyperref}       
\usepackage{url}            
\usepackage{booktabs}       
\usepackage{amsfonts}       
\usepackage{nicefrac}       
\usepackage{microtype}      
\usepackage{lipsum}
\usepackage{graphicx}
\usepackage{enumitem}

\graphicspath{ {./fig/} }

\title{Open Challenges of Blind People using Smartphones}

\author{    
  Andr\'e Rodrigues\textsuperscript{1}\thanks{Corresponding Author}   ,  Hugo Nicolau\textsuperscript{2}, Kyle Montague\textsuperscript{3}, Jo\~ao Guerreiro\textsuperscript{2}, Tiago Guerreiro\textsuperscript{1} \\
  \textsuperscript{1} LASIGE, Faculdade de Ciências, Universidade de Lisboa, Portugal\\
  \textsuperscript{2} INESC-ID, Instituto Superior T\'ecnico, Universidade de Lisboa \\
   \textsuperscript{3} Open Lab, Newcastle University, United Kingdom \\
  \texttt{afrodrigues@fc.ul.pt, hman@inesc-id.pt, kyle.montague@ncl.ac.uk, jpvguerreiro@gmail.com,}
  \\
  \texttt{tjvg@di.fc.ul.pt} \\
}

\begin{document}
\maketitle
\begin{abstract}
Blind people face significant challenges when using smartphones. The focus on improving non-visual mobile accessibility has been at the level of touchscreen access. Our research investigates the challenges faced by blind people in their everyday usage of mobile phones. In this paper, we present a set of studies performed with the target population, novices and experts, using a variety of methods, targeted at identifying and verifying challenges; and coping mechanisms. Through a multiple methods approach we identify and validate challenges locally with a diverse set of user expertise and devices, and at scale through the analyses of the largest Android and iOS dedicate forums for blind people. We contribute with a prioritized corpus of smartphone challenges for blind people, and a discussion on a set of directions for future research that tackle the open and often overlooked challenges. \end{abstract}


\keywords{.}

\section{Introduction}
The emergence of smartphones confronted their users with revolutionary changes in the way these are interacted with. The changes go beyond the lack of physical keys and boundaries: the number and diversity of applications exploded, several applications can be used at the same time and inter connectedly, user interactions can vary depending on context, among others. Alongside, these personal devices are the ones being used by the largest amount of people in the world, and their usages have long surpassed communication, being everyday tools of productivity and leisure. 

On one hand, smartphone capabilities have brought forward a new set of opportunities for the creation of ubiquitous assistive tecnhologies. Developers are now able to create a variety of services that were previously only available through specialized expensive equipment. Nowadays, through installable apps, users can take advantage of smartphone devices to take photos and ask questions of volunteers (e.g. \cite{bemyeye, Bigham10}), automatically identify colors (e.g. \cite{colorid}), read documents using optical character recognition (e.g. \cite{knfbreader}), add audio labels to items through QR codes stickers (e.g. \cite{DigitEyes}), among many others. Besides the paraphernalia of applications that came along with these new devices, another contributing factor to adopt them is social acceptance. Feature phones became obsolete and, despite their recognized higher accessibility, became socially undesirable \cite{Shino2011,Kane09,Pal17}.

Conversely, the paradigm shift in mobile device user interface design also brought up new challenges for accessibility, and particularly so for blind people. The increasing relevance of these devices placed the spotlight on research and development around the accessibility of touch-based smartphones. Still, this research has been characterized by a superficial lens, or even stereotypical, over non-visual interaction with touch surfaces: focus has been mainly given to gestures on a touchscreen and the text-entry task, as an extreme demanding example of touch interaction with a grid of small targets. While this first barrier is relevant to enable access to the device, the challenges imposed by the new multi-application environment and its chamaleonic user interfaces are not of lesser relevance nor impact. These  may not be at the surface nor arise in a controlled environment given that is not the context in which mobile interactions take place.

Our approach has been to seek a more in-depth understanding of the problems faced by blind people in their day to day interactions with mobile devices. In a first study [\cite{Rodrigues15}], we investigated the challenges, needs, and coping strategies of 5 novice visually impaired users over a 8-week period, following a mixed-methods approach. These have brought up severe challenges in adopting the change of paradigm, from feature phones to smartphones, and the continuous over reliance on third-parties, one that limits independence and autonomy. Results corroborate that the touchscreen interaction was an issue but that other interaction challenges severely impair usage. 

In this paper, we go further in understanding the everyday challenges faced by blind people when using smartphones, during their first encounters with these devices, as well as those that continue to be issues after adoption. We present and discuss a set of 3 studies performed with blind people of different mobile usage expertises, focusing on their current and past experiences with mobile phones: 1) a series of workshops with expert and novice users to understand barriers found, surpassed, and coping strategies (summarily presented in \cite{Rodrigues17}); 2) a thematic analysis of questions and doubts on online forums for accessible technology; and 3) a card ranking study to prioritize the corpus of challenges collected. 

We contribute with: (1) a characterization and a corroboration of the mobile challenges faced at different expertise levels; (2) an understanding of the coping mechanisms employed; (3) an assessment of the severity of challenges based on expertise level; (4) a reflection on the use of multiple methods in accessibility research; (5) and,  a discussion on the open challenges characterized to provoke future research in mobile accessibility beyond stereotypes.

\section{Related Work}\label{class}

In this section, we outline prior work on touchscreen accessibility for blind and visually impaired users identifying the existing knowledge gap in the broader understanding of smartphone accessibility. We discuss prior work that tackles accessibility issues in smartphones from substituting the entire system to adapting content. Our work is informed by how past work in accessibility research has relied in a multitude of methods to overcome the limited views and verify what spawned from controlled laboratory assessments.

\subsection{Touchscreen Accessibility for Non-Visual Input}

Smartphones are becoming the norm and touchscreens are the current mainstream input method that accompanies them. As such, researchers took a particular interest in first understanding gestural interaction for blind people \cite{McGookin08,Kane08,Kane09,Oh13,Buzzi17}. Nowadays, smartphones come preloaded with a wide range of accessibility features, some of which are designed specifically for non-visual interactions. Android's Talkback and iOS' VoiceOver are screenreaders that enable visually impaired users to navigate the entire system with a set of defined gestures. Users can swipe to navigate sequentially a list or grid of elements; scan the elements in the screen with their finger, where the system reads aloud the current focus; or double tap to select the current element - similar to the techniques proposed by \cite{Kane08}. While these accessibility services allow blind users to interact with smartphones, they can result in much slower interactions than those of sighted users \cite{Oliveira11}.

Text-entry on touchscreens is one of the most researched topics. Several studies focused in understanding text-entry needs on touchscreens \cite{NicolauTypingPerfomance15,Rodrigues16} while others looked for alternative interaction methods, many of which based on Braille \cite{AzenkotWobbrock12,Nicolau15,NicolauMontague14,OliveiraBrailleType11,Romero11,Southern12}. Difficulties in operating a touch-based phone without visual feedback have also been acknowledged by other researchers working on haptic feedback \cite{Buzzi13} or even on how to teach touch gestures to blind people \cite{Oh13}. 

While non-visual operation of a touch-based smartphone seems to be a concern, there are, to our knowledge, no reports on the very first contact and daily usage of these devices by blind users. Moreover, we have no knowledge of the daily challenges users of different expertise levels are facing. Research focuses on the stereotypical challenges such as text-entry and gestures, which can potentially be a small part of a blind users' daily challenges. 

\subsection{Adapting Smartphones}
Although unexplored, researchers and developers are aware that smartphones still have accessibility issues. Applications suites such as Mobile Accessibility  \cite{mobileaccessibility} or EasyPhone \cite{easyphone} were developed  in the earlier days of smartphones, specifically to facilitate smartphone usage for visually impaired people.  These approaches replace the entire system and limit the user to a set of provided applications, guaranteeing accessibility but restricting access. Since smartphones started to come with better accessibility options and compliance out of the box these approaches have become outdated. 
Rather than substitute the entire system one can adapt how interfaces are displayed, keeping the original content and possibly enriching it with other sources \cite{RodriguesStatic17, Zhang17}. Zhang et al. \cite{Zhang17} explored this concept and argued for the  use of interaction proxies that enabled third party developers to be the ones to address apps' accessibility issues. Unfortunately, the premise of relying on outside developers to address app issues means they have to not only to fix issues, but also ensure their adaptations do not compromise any features. Moreover, it relies on a particular limit set of developers who are accessibility knowledgeable. 

\cite{RodriguesStatic17}  divided the screen in two halves, dedicating the top half to a set of fixed options emulating feature phones with tactile keyboards, allowing the reordering of content. The adaptation reduced the complexity of interacting with a smartphone and was seen by participants as a stepping stone in the learning process since it hid many of the expert features present in modern mobile screenreaders.

All the solutions described rely on the understanding of the challenges and coping mechanisms blind people have when interacting with smartphones. While some challenges have been identified in prior work, as reported in the surveys presented in \cite{Damaceno2018,Grussenmeyer17}, so has the need for a real-world assessment of the usage of mobile technologies by blind users \cite{Grussenmeyer17}.

\subsection{Accessibility Research}
Prior research has investigated technology adoption and usage by people with disabilities in different contexts. Through interviews and participant diaries, \cite{Kane09} assessed the challenges faced by people with visual and motor impairments when using mobile devices. Similarly  \cite{Pal17}  used surveys and interviews to ask blind smartphone users what their devices enabled them to do and how they were taking advantage of these abilities. \cite{Pal17} went on to argue for the need to understand human agency within the necessity of adopting technology. The previously referred research provided insights about real-world technology usage by individuals with disabilities such as the first contact with AT technologies in low-mid incoming countries primarily happens in non-profit organizations. However,  data is limited to a particular time-window, dependent on self-reporting and on the reflective capabilities of participants to elaborate on their experiences. Moreover, when we take into consideration that one of the challenges blind people face when interacting with content is "not knowing what they do not know'' \cite{Bigham17} we realize the need to go beyond reflective approaches.

Through YouTube videos of people with physical disabilities interacting with touchscreen devices and supported by online surveys, \cite{Anthony13} assessed how this user group was engaging with technology. Thus, capturing the unique and interesting ways in which people have augmented or crafted solutions to support their interactions with touchscreen devices. Furthermore, the study highlights real-world usage scenarios and interaction contexts. Similarly, \cite{Branham15} conducted contextualized, semi-structured interviews in participants' homes to understand the challenges to creating accessible shared home spaces between sighted and blind co-habitants. \cite{Naftali14} conducted four in-the-wild case studies with people with motor impairments exploring the impact of environmental context on their mobile interactions using a combination of interviews, participant diaries, and contextual session observations.

While laboratory studies may not be able to uncover or verify real world problems, typically, in-the-wild user studies are not able to obtain objective performance measurements of device interactions. One exception is the work by \cite{Montague14} that by relying on a custom built game had the means to capture touchscreen interaction performance. However, such approaches are limited to collecting data within the bespoke applications - significantly overlooking the interactions that participants were making with other applications (where more time is spent).  Moreover, in-the-wild studies that rely on a single methodology (i.e. interviews; observations; or device data collection) result in an isolated understanding or limited view of the challenges. None of the referenced research works is of significant length or relies on a variety of data sources to inform their findings, thus one can always questions its applicability.  Each study, methodology and data source has its own strengths that in isolation  may provide an incomplete representation of the barriers and have a narrower claim to the validity of its claims.

To address these issues, one can rely on a mixed-methods approach where in a single study many different data sources are collected (e.g. interviews, usage and performance data) that together can aid the sensemaking process of the findings. Alternatively or in cooperation, one can also rely on a multi-method devising several studies investigating the desired phenomenon through several lenses.

\section{Methodological Framework}
In this work, we go  beyond the state of the art by leveraging multiple research methods to explore, characterize and verify the challenges faced by blind people.

In our past work, we sought to understand the adoption process of newcomers \cite{Rodrigues15}. To do so, we conducted a 12-week in-the-wild longitudinal study with five blind users. The study relied on a mixed-method approach with pre-adoption and weekly interviews, weekly task assessments within a controlled environment, and data collection of participants in-the wild smartphone usage (system-wide). The aforementioned study was the genesis of the work presented in this paper. The small scale study unveiled participants' concerns, barriers faced, support mechanisms and usage evolution. The study prompt us to question: 1) how and if long term users are still dealing with barriers, 2) do the findings apply at a larger scale with a variety of devices and operating systems, and 3) what is the relative importance of each challenge and what can be done about it.

To understand the extent of the challenges found across a larger user base and expertise level, in the first study presented in this paper, we conducted a series of workshops locally in Portugal. We report on the challenges and coping mechanisms of 40 blind screen reader users. A summary of this study has been previously presented in \cite{Rodrigues17}.

To capture the challenges, without any researcher intervention, and understand how solutions are provided to a larger and more diverse population, we analysed discussions from the largest English speaking Android and iOS dedicated forums for blind people. In contrast with our previous studies, where we had a limited number of users who had fully adopted the device and all its features, challenges discussed by the forum users suggest a high proficiency with the smartphone devices - especially from those providing answers and support. 

Through the aforementioned studies, we identified and verified multiple causes for the difficulties blind people face when interacting with smartphones. 

To understand the importance that blind users would assign to the various concerns and challenges, as well as the existence of conflicting attitudes that might arise from the diversity in users' experience level, we conducted a card ranking study.  In this study, we assess the severity of challenges and concerns based on expertise level.

\section{Study 1: Newcomers, Novices and Experts Workshops}
In this study we sought to verify if the problems previously reported by \cite{Rodrigues15} were representative of the challenges faced. Moreover, if so, we were interested in understanding  how they were being addressed. Thus, we recruited  blind screen reader users, with different expertise and devices. In an effort to observe what challenges naturally occur during use, and understand what and how people cope, we conducted a series of semi-structured workshops where participants  led part of the session based on their questions.

\subsection{Participants}
We had a total of 42 blind participants, 23 males and 19 females, with ages ranging from 25 to 79 (M=51.8, SD=14.0) years old. Participants were recruited through social media, word of mouth and through a local social institution for blind people. Users were asked about their smartphone proficiency when registering for the workshop:  17 did not own any smartphone nor had any previous experience with it; 18 had a device but considered themselves novice as they were only able to perform simple tasks; and, the remaining eight considered themselves experts.   

\subsection{Apparatus}
The workshops were conducted in a room provided by the local institution. All participants were asked to bring their smartphones to use during the workshop, resulting in a wide variety of brands and models. We also provided a smartphone to all participants that did not own one. The Android devices used the default Talkback screen reader while iOS devices relied on VoiceOver. Sessions were video and audio recorded. 
\begin{table}[]
\begin{center}
 \begin{tabular}{||l| l| l||} 
 \hline
 Session & Participants & Nº Sessions\\ [0.5ex] 
 \hline\hline
 Android Novice & 26 & 3 \\ 
 \hline
 Android Expert & 1 & 1 \\ 
 \hline
 iOS Novice & 9 & 2 \\ 
 \hline
 iOS Expert & 6 & 1\\ 
 \hline
 \end{tabular}
\end{center}
\caption{Workshop sessions conducted.}
\label{table:3}
\end{table}
\subsection{Procedure}
At the beginning of the sessions participants completed a short questionnaire related to their smartphone usage and general demographics. We gave two types of workshop, one for the newcomers and novice users and the second for experts. During the sessions we engaged participants in a group discussion. These sessions had a duration between one hour and a half and three hours depending on group size and questions. Participants guided most of the session as they were free to ask questions and collaborate during the sessions. We had separate sessions for Android and iOS with a total of seven sessions (Table \ref{table:3}). For all sessions we had a team of four researchers available with experience in  mobile devices accessibility, who assisted the sessions always guaranteeing a ratio of 2-3 participants per researcher.  Each researcher was also responsible for taking notes and observations.

\subsubsection{Session for Newcomers and Novice Users}
The session was divided in two major parts. First participants were guided through basic smartphone and screen reader behaviours, then participants took the lead and were supported individually or in groups depending on their questions or what they desired to learn.

In the first part, participants started by learning about smartphones and the differences to feature phones. Afterwards, participants were guided on how to perform gestures and basic tasks (e.g. navigating the screen, using the contacts app, writing text). In the first session we conducted, users completed only the first two lessons,\textit{ Exploring the Screen and Using Lists}, since in \cite{Rodrigues15} we found the last two lessons of the Talkback tutorial to be too demanding on newcomers. During the session, it was clear that participants were struggling even with the first two lessons. Thus, we dropped the tutorial and we gave personalized instructions to each participant. 

Although this was the basic structure for the workshop, as the level of expertise varied greatly amongst newcomers and novice users, some led the session from the beginning skipping the first part, and learned about more advanced features (e.g. adding a contact or deleting messages). In the iOS workshop, participants were not newcomers and could already perform some of the basic tasks (e.g.call a contact) we started the session by introducing the VoiceOver Practice. Participants could freely perform gestures and when a gesture was recognized (e.g. "Touch - Select item under your finger") its function was read aloud. The VoiceOver Practice allowed us to introduce and explain gestures they were not familiar or struggled with. Afterwards participants led the session and learned to perform tasks they were not familiar with but wanted to learn(e.g. add a contact).

\subsubsection{Session for Expert Users}
We had two expert sessions, one with a single Android user and a second one with six iOS users. Participants were expected to come with doubts regarding the usage of their device, thus sessions were fully led by participants interest. The Android session was a one on one session since we only had a participant. In the iOS session, we engaged the six participants in a focus group discussion. Participants exposed questions and anyone could contribute  based on their experiences. This approach allowed us to observe how participants convey their knowledge when co-located. Moreover due to the range and specificity of the questions, input from other expert users proven to be crucial to understand the underlying issues and be able to quickly address them. After all questions were answered, we inquired what types of barriers they face during usage, how they tackle them, how they learned to use the device as newcomers, and if/how they assist others. 

\subsection{Findings}
We report on the qualitative data collected during the sessions through the researchers perspectives. We conducted a thematic analysis using the protocol suggested in \cite{Braun06}. We first start by identifying the basic codes on the researchers' notes of the workshops, and progressively iterated and discussed the emerging themes with the four researchers that participated in the sessions. When the noted observations referred to participants comments we relied on the audio recordings to transcribed said comments. When behaviours were described that required further assessment we relied on the video recordings.

\subsubsection{Getting Started}
Prior to the workshop, some of the participants reported they had tried to use smartphones only to give up on the process for being too cumbersome.  The current mechanisms to support blind newcomers on mobile devices are insufficient. Although, Talkback on Android provides a tutorial to get started, it frustrates and misguides users as we had reported previously \cite{Rodrigues15}. In the iOS session users started in the VoiceOver Practice which enabled them to understand how their actions were being interpreted by the gesture recognizer. However, by design the VoiceOver Training requires users to already be aware of the gestures available or have someone tell them to try specific gestures.  We observed the same problems reported in \cite{Rodrigues15}, smartphone adoption challenges do not appear to be  device or operating system dependent. Currently picking up a smartphone and start using it without assistance is a difficult task with little to no native support. 

\subsubsection{Discoverability}
Only one iPhone participant was a newcomer, yet none of the novice participants was aware of some of the basic gestures they could perform on the device, such as flick to the next item. Most relied exclusively on tapping (i.e. instead of flicking or dragging their finger through) the screen until they eventually found the option they were looking for. Participants thoroughly discussed issues they experienced with smartphones. For newcomers, their problems were related with touchscreen interactions and simple gestures. However, the cause of their struggles was related to a lack of understanding on how the underlying interfaces were behaving.  Participants reported that often they were unaware of the available options ("\textit{In one app I had no way of sharing to Facebook. When I pressed More Actions nothing happened. What I found out afterwards, when I asked a friend, was that the option was there but it was not yet on the screen. I had to scroll on a new window that appeared.}"). All participants reported issues with smartphones, independently of expertise level and device. However, expert users focused more on application-specific issues, such  using advanced features ("\textit{I am not able to listen to music from my Dropbox in offline mode}").  We observed the same behaviours in Android users, discoverability challenges do not appear to be device or operating system dependent and are one of the main challenges users seem to face.

\subsubsection{Independent and Community Learners}
Participants strongly rely on others to surpass challenges, often asking for help from people they consider to be technology experts. We found that users informally created communities that relied on the same specialist; two of them were present in our workshops. They were tech savvy, autodidact, and highly motivated to learn about technology. They regularly read blogs, forums, and mailing lists about assistive technologies, and even contact developers to report bugs and request features. Several participants in the workshop relied on them to cope with daily problems. They provided assistance through a variety of channels (e.g. calls, SMS, Skype) and often about the same issue but to different people.  They reported how their expected availability given their role in the community as strain their more intimate and personal relationships. 

\subsubsection{Sighted Assistance}
For some issues, the only possible solution was asking for help from a sighted friend (e.g. screen reader started speaking in a foreign language). However, participants discussed some situations where help from sighted friends and family was challenging due to their unfamiliarity with screen readers. All but one participant mentioned how they preferred to be helped by screen reader users. \textit{
"Often the problem is not them [sighted users] not knowing how to solve the problem, the problem is not knowing how to explain to us how we can solve it"}. 

Although sighted people are seen as valuable sources of assistance, most of them are oblivious to the challenges of screenreader users. They usually know the steps needed to accomplish a given task, but are unaware on how to perform them using accessibility services.

\subsubsection{Detailing Instructions}
During the workshops, experienced users would often help by guiding others step-by-step, while doing the actions on their own devices and waiting for others to finish each step. For gestural interaction, some participants went further and performed the gesture on the back of the other users' hand. Nevertheless, it was clear that people preferred an active learning approach rather than giving their device to others. 

The level of detail given can be fundamental to a successful  instruction. \textit{About the VoiceOver rotor gesture is "To use the rotor, rotate two fingers on your iOS device's screen as if you're turning a dial. VoiceOver will say the first rotor option. Keep rotating your fingers to hear more options. Lift your fingers to choose an option."} Although this is an already rich description one of our participants mentioned how it was not enough:
\textit{"Two fingers? Which fingers? Rotate to what side? Rotate how exactly?"} The participant carried on and mentioned how he learned that day how to effortlessly perform the gesture when one of the researchers described it to him \textit{"Use your thumb on the screen and then with your index finger you are able to rotate similarly to a compass when you hear the option you want to select lift your finger".  }Before the user heard this description he was using his index and middle finger to rotate at the same time and was unable to use it. The researcher was only able to provide this instruction due to the participants previous remark. 

\subsection{Discussion}
One lingering question from the adoption study of \cite{Rodrigues15} was whether problems existed at different levels of expertise and with different devices. In this workshop we observed 40 blind people with fifteen unique devices, two different operating systems, and different levels of expertise; the observations, comments, and questions suggested that although barriers can become easier to surpass, they continue to exist. With this study we verified at a local level the issues identified previously are pervasive in the community. 

\subsubsection{The challenges of awareness, creating mental models, and accessibility issues}
Challenges are more often than not related to a lack of awareness of the surrounding options triggered by the difficulties in establishing a complete mental model. When starting to use smartphone apps, participants struggled to find consistency among and within applications. Applications no longer have a single sequence of steps to move from point A to point B,  but several alternatives to achieve the same goal. They can also have workflows loops that may be difficult to recognize. Some of the problems with understanding patterns and workflows resemble the ones reported in web accessibility with inter and intra page navigation \cite{vigo13}.  However, traditional solutions may not work for smartphones:  screens don't have unique links or any identifiers, navigational breadcrumbs are not used or practical given the small screen real state, and back functions are not standardized across apps or OS.   Moreover, applications are created by different developers, which makes it even harder to find consistency among apps interfaces. The issue is exacerbated  as users explore more of the device and third party applications. It becomes highly likely they are confronted with inaccessible content, from elements hidden to navigation, to unlabeled ones. Additionally, although we found no instances of this issue during the workshops,  mental models may be disrupted by app and OS updates.

\subsubsection{The need for assistance does not disappear}
Different devices come with different characteristics but the fundamental problems remain. The support mechanisms on smartphones are not enough and to surpass the challenges users often resort to external help. Unfortunately, others are not always available and online help (i.e. checking forums and reading tutorials) is only reported to be used by a few tech savvy experts. Furthermore, our findings suggest this is causing issues beyond smartphone accessibility, with newcomers abandoning devices and experts having their relationships strained due to tech assistance.

\section{Study 2: Community Forums}
In the expert sessions of the previous study, we observed that the experts who were providing assistance within the community were  also the ones that  rely on technology like forums and online guides to overcome issues. Likewise, many others around the globe do the same, thus to capture a wider range of problems and understand how solutions are presented in a environment without any researcher intervention,  we analysed the top content in the largest Android and iOS dedicated forums for blind people. The data collected allow us to understand how knowledge is shared in-the-wild in an online community. 

\subsection{Data and Analysis}
In this study we started by selecting one forum of both major mobile operating systems (OS). Secondly, we analysed the top thread titles and selected threads with content relating to barriers and assistance for further inspection. We then performed a thematic analysis on the selected content with a focus on understanding the type of barriers and how solutions are provided. 

\subsubsection{Dataset}
Although there are multiple relevant online communities in both OS, we selected the ones with the higher number of content. \cite{viphone}, the iOS representative, is a Google group with over 41989 threads. \cite{eyesfree} is the Android Google group with over 27898 threads. Google groups is a service from Google that allows the creation of discussion groups. The content can be created and consulted online in a forum based interface or as an email based system. When relying on an email client, the number of views are not counted towards the thread statistics. Thus, we used the sum of the number posts and views in each thread to select the highest ranking content. To discard outdated discussions we limited our search to threads with content from January 2015 onward. One researcher coded the title of the top 100 threads of each forum. All titles coded as \textit{Doubt} (i.e. question about something), \textit{Problem} (i.e. specifies an issue), \textit{Guide} (i.e. app, feature or device guide), \textit{Getting Started} (i.e. mention getting started) and \textit{Request} (i.e. request for information) were selected for further inspection. Then we analysed the first message and discarded all threads that did not discuss smartphone applications, devices or features, resulting in 48 selected threads (i.e. 19 from Android and 29 from iOS). The majority of the discarded threads were coded as \textit{Announcements} (i.e. release announcement app/product/device). The selected threads have 45 unique authors, a total of 2502 posts and 10968 views. Posts per thread range from 2 to 660 (M=52.12 SD=100.98) while number of views from 9 to 4173 (M=224.14, SD=601.29).

\subsubsection{Data Analysis}
We conducted a thematic analysis for the 48 threads. We used a combination of inductive and deductive coding; two researchers coded a set of 10 threads independently. Then the two researchers discussed, iterated and merged the two codebook. The final codebook was used by one researcher to analyzed the remaining 38 selected threads. Threads  had as many as 660 messages and often after a number of posts the discussion either shifted to irrelevant content or repeated arguments. Thus, to prevent analyzing unrelated content we stopped coding a thread if three sequential posts were marked as Other content. To reduce the amount of repeated content after the first 10 posts we started assessing if the next three posts contained novel content, if not the analyses of that thread was stopped. We coded a total of 524 individual posts (M=10.91, SD=6.47 posts per thread).

\subsection{Findings}
In this section, we present the two overarching themes that led the data exploration: \textit{Barriers and Solutions}. In Barriers we identified the challenges described by users; in Solutions we report which and how they are provided. In this section we use "users" when  referring to the forum users and "authors" to indicate thread authors.

\subsubsection{Barriers}
Barriers are depicted in 41 threads (85\%). The remaining threads are open discussions where authors request other users opinions about a particular device, feature or application. 

\paragraph{\textbf{Text-entry}}
As expected, we found multiple instances on the top relevant content discussing text-entry problems. Users are slower entering text in touchscreens when compared with physical keyboards. Consequently, users look for alternative means to write from QWERTY to braille virtual keyboards or physical keyboards.
\textit{"iOS braille is so much faster for me and I get much less errors. "
}
However, speed is not the only challenge users face. Currently, editing, copying and pasting text is cumbersome. Users wished they found a simple way to enable them to manipulate text as they are accustomed to in other systems.

\textit{"What bothers me is the clunkiness of the current system for selecting and editing text." }

\paragraph{\textbf{Gestures}}
In this analysis we did not find any gesture related issues with the exception of the required gesture to answer a call. The underlying challenge was not how to perform the gesture but rather know how to interact with the interface presented. However, many of the described issues in other threads, may be caused by gesture difficulties that the user is unaware of. For example, not finding a particular element on the screen, due to an incomplete exploration, may be caused by users skipping elements during exploration. Suggesting the issues described in the first study are relevant beyond the local environment.

\paragraph{\textbf{Visually rich interfaces}}
A rich visual interface allows a sighted user to quickly understand the structure of the content. On the other hand, visually impaired users have to scan through all of the interface to create a mental model of it. Moreover, the navigation using a screen reader is affected by the underlying interface grouping. Therefore, for an efficient navigation users must create a mental model that incorporates the interface structure in each application. 

\textit{"Remember! If you move your finger to the top of the screen, into a different area of the screen, you will "automatically!" go back to the default navigation/group level. So, do keep that in mind."}

Structure and correct labeling of elements is essential for the accessibility of an app, but it does not guarantee it. There is information that is lost when using a screen reader. When first playing Mine Sweeper, a sighted user will quickly understand the rules and objective of the game through the visual cues provided. In a fully accessible Mine Sweeper application a user that is unfamiliar with the game will struggle to understand it.
\textit{
"I've never played a mine-sweeper game before and so, when playing their accessible mine sweeper, I didn't have a clue what to do"}

\paragraph{\textbf{Discoverability}}
Applications features need to be clear and discoverable. We found multiple reports of users unable to perform an action in an application, not due to content that is impossible to access but due to a lack of discoverability, corroborating the first study findings. 

\textit{"Somehow the Alarm volume got set at a very, very high loudness level. I can't seem to find a way to tone it down."}

\paragraph{\textbf{Dynamic content}}
In this study we found that one of the reasons users struggle to create accurate mental models is because many mobile applications have dynamic content. Apps where repetitive interactions cause different outcomes disrupt users mental models, similar to what happens when the back button behaves inconsistently \cite{Rodrigues15}. A common case of this behaviour is when users are faced with an ad during their interaction.

\textit{"Then, without me appearing to do anything, I seemed to be hurled, into the world of vehicle and personal accident insurance.  Where in the wide world that came from, I haven't the foggiest."}

\paragraph{\textbf{Efficiency}}
Smartphone interfaces are designed to be visually appealing and provided fast direct manipulation. Some tasks are inherently associated with the underlying interface structure and item location. When using a system-wide screen reader that modifies every interaction, these tasks may become cumbersome or even impossible to perform.

\textit{"I can go into the app drawer and tap and hold on an app to add it to the home screen, but this seems an awfully long way of doing things and I still can't add a widget or shortcut which is pretty frustrating.  So what am I doing wrong here? I'm so confused and most of all frustrated."}

\paragraph{\textbf{Sighted Assistance}}
As described in the previous study, non-screen reader users may not be able to use the device due to its different interaction method. Thus, it can prevent them from providing effective assistance.  

\textit{"In my experience, even just having VoiceOver on makes the phone fairly difficult to use for a sighted person, since the VoiceOver gestures are very different from the way a sighted person would normally interact with a touch screen device." }

\paragraph{\textbf{Updates and Versions}}
In this study we confirmed what we posited could further disrupt users mental models and prevent users to take advantage of all features. Smartphone OS and applications are updated frequently, which can lead to user segregation with respect to the versions available or installed on devices. Individual versions may possess different features and requirements thus, it can restrict the user's access. 

\textit{"They don't let me buy it!!! Probably, Google sees my Galaxy s3 as an "old Samsung device", not knowing it is now running 5.1.1... "
}
Updates often bring new features and behaviours. Unfortunately, more often than not the changes also cause new bugs and barriers to appear. When there are screen reader or  OS updates it potentially affects the whole user experience with the smartphone.

\textit{"I've upgraded to 5.0.2 in my Moto G 2nd gen, and now, I'm facing audio cut offs. Talkback is not speaking properly, and audio quality has also been decreased. "}

\subsubsection{Solutions}
In every thread where a barrier was presented other users contributed to the discussion by sharing their experience, providing solutions, and clarifying features. However, not all problems were solved. In some cases all the authors was advised to do was to contact the developer or manufacturer due to the nature of the barrier. In this section we report on the different types of solutions and how they were presented to users. 

\paragraph{\textbf{Proficiency dependent instructions}}
When a user had the knowledge to solve a problem they would often guide the author to the solution. When providing the answer the user would assume a certain degree of proficiency of the author. If the author was experienced, the instructions given could be presented at a high level just describing the overall steps needed to be taken.

\textit{"It is called NuPlayer, and you can disable it in the developer settings."}

In cases where the author was inexperience or asked for further guidance each step was presented in greater detail including the required gestures/navigation and received feedback.

\textit{"OK, first you must enable developer options. Go to Settings, About Phone, then tap on the build number about seven times. (...)"}

\paragraph{\textbf{Location information}}
Unexpectedly, most steps do not contain the relative or absolute position of the items. Users usually are simply instructed to find the item \textless x\textgreater
 in screen/layout \textless y\textgreater. Of the 82 references to guiding, only 12 contained item location reference. When location is provided it is usually absolute or relative to the always present keys (e.g. home, volume). Absolute locations are provided to items close to the edges of the device (e.g. bottom left, top right). We only found one instance of an absolute location not associated with an edge (i.e. \textit{"check in the middle of the screen, toward the top"}) and one of relative location associated with a virtual button (i.e. \textit{"find the 'audio switch' on right of the ‘dial pad switch'}").  

\paragraph{\textbf{Sighted People Required}}
Unlike our previous studies, we only found two instances where sighted assistance was required. In a particular case, a user was inquiring about a phone and it's out of the box experience. Unfortunately, that particular device required sighted assistance to enable the accessibility options. 

\textit{"The out of the box experience, is not quite so good though, as you can't independently set it up."
}
The second instance was an unexpected use for a sighted assistance. Users asked the author if he could reach out to a sighted person so he could provide greater detail about the problem he was facing.

\textit{"If you have enough vision or have a sighted person available, can you let us know if the VoiceOver visual cursor is also skipping various elements as well?"}

\subsection{Discussion}
We analyzed the content of two open online communities dedicated to smartphone accessibility. Therefore, we do not have any demographics or information about each individual. However, we know users are able to communicate in English, and that when faced with a  barrier relied on online communities to overcome it. Even without assessing the content of each request these users are at the very least proficient enough with either smartphones or computers to browse and query for questions online. In contrast, in our previous studies, users reported the need for co-located assistance, the lack of its availability and their high dependency on others. In this study, the majority of users tried to address the topic by discussing possible solutions and by guiding the thread author to a successful outcome without external help. Only when the problem seemed to be unmanageable did users suggest to get outside assistance (e.g. retail store, co-located sighted assistance). 

\subsubsection{Instructions are concise, neglect location and app structure}
Users provided concise instructions, guiding the user with high level instructions when possible. When more detailed instructions were required, either by author's request or due to an assumption on the author's expertise,  users provided step-by-step instruction providing the target element description. Although, many of the problems seem to stem from the lack of understanding of the underlying app structure or how to reach elements efficiently, instructions do not contain location nor an overview of the app layout. This can be accredit the forum being dedicated to visually impaired users, thus the visual representation and element location can be hard, if not impossible, to understand. Moreover, there are two diverse navigation methods that each user can rely on, thus what would be a helpful app overview for a linear navigation might not be for an explore by touch approach.

\subsubsection{Lost in constant updates and dynamic content}
Many of today's mobile apps rely on dynamic content for a variety of reasons, from add revenue to keeping the user updated with the latest news. However, as many changes in context and content are strictly announced through visual stimuli making their underlying cause and sometimes even context undetectable for blind users, other than realizing something changed on the screen. Thus, users often feel a loss of control that can even be disorientating. As a side-effect it can cause users to struggle to create accurate mental models. Several of the most popular threads analysed were discussions around the impact, bugs and issues of the updates to either applications or operating systems. These  constant updates and revisions to both function and interface only escalate the difficulty of creating mental models which is already a demanding task to accomplish.

\begin{figure}
\includegraphics[width=\textwidth]{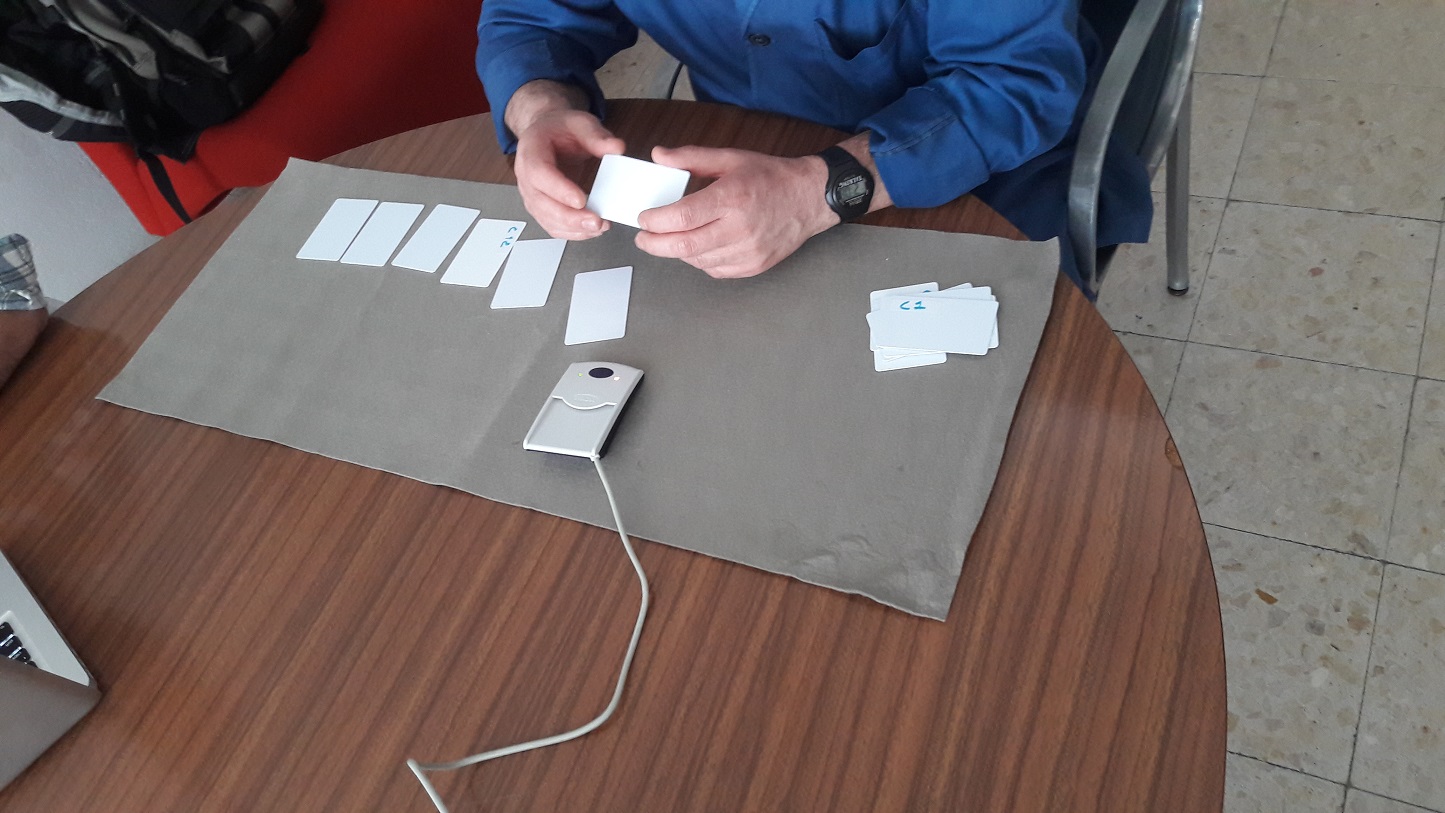}
\caption{User ranking a RFID Card.}
\centering
\label{figure:1}
\end{figure}

\section{Study 3: Prioritizing Challenges}
In this study, we aimed at understanding the importance that blind users assign to the various concerns and challenges we derived from our user research, as well as the existence of conflicting attitudes that might arise from differences in users' experience level or device.

We sought to contribute to existing literature in two ways. First, to the the best of our knowledge, the present investigation is the first attempt to prioritize smartphone challenges based on perceived level of severity. Second, this study is the first attempt to use rank-ordering with blind users as a participatory design activity to promote communication about relative perceived importance of smartphone challenges.

We examined three main research questions: 1) Among 13 specific challenges, what was the order of perceived importance by blind users? 2) Does experience have an effect in the perceived importance for each of the 13 challenges? 3) What were the participants' views on the rank-ordering activity?

\begin{table}[]
\centering
\begin{tabular}{@{}llllll@{}}
\toprule
Participant & Gender & Age & Device  & Time      & Expertise \\ \midrule
P1          & M      & 58  & Android & 24 months & Novice    \\
P2          & M      & 63  & Android & 18 months & Novice    \\
P3          & M      & 38  & iOS     & 48 months & Expert    \\ 
P4          & M      & 44  & iOS     & 48 months & Expert    \\
P5          & M      & 57  & iOS     & 14 months & Novice    \\
P6          & M      & 37  & Android & 48 months & Expert    \\
P7          & M      & 27  & Android & 48 months & Novice    \\
P8          & M      & 49  & iOS     & 5 months  & Expert    \\
P9          & M      & 39  & Android & 48 months & Expert    \\
P10         & M      & 26  & iOS     & 48 months & Expert    \\
P11         & F      & 35  & Android & 18 months & Novice    \\
P12         & M      & 41  & iOS     & 54 months & Expert    \\
P13         & F      & 57  & Android & 3 months  & Novice    \\
P14         & M      & 63  & Android & 7 months  & Novice    \\
P15         & F      & 27  & iOS     & 3 months  & Expert    \\
P16         & F      & 58  & iOS     & 6 months  & Novice   \\ \midrule 
\end{tabular}
\caption{Participants profile.}
\label{table:4}
\end{table}

\subsection{Participants}
We recruited 16 blind participants (4 females) from a local training center. Half of the participants used an Android device (8 participants), while the other half used an iOS device. Participants used a screen reader to access their mobile devices and their ages range from 26 to 63 (M=45, SD=13). Our final user sample consisted of participants with 3 to 54 months of smartphone experience. Because we aimed to recruit a sample of participants with different experience levels, we asked what tasks they were able to perform with their smartphone, from making voice calls, send text messages and email to listen to music, install new apps, and web browsing. Rather than use time as a measure of expertise, we considered the amount of tasks participants were able to accomplish with their devices. We considered those who were able to perform all suggested tasks as expert users or novice otherwise. Although some novice participants may actually be able to perform more advanced operations, they were still not able to fully control their device. For instance, P1 and P5 read emails frequently, but were never able to compose a new email. On the other hand, P8 only used  a smartphone for 5 months and was already able to perform all suggested tasks. Table \ref{table:4} shows participants' profile.

\subsection{Methodology}
Based on the results from the previous studies and an analysis of the literature (\cite{Oh13,Rodrigues15,Nicolau17,azenkot13,Bigham17}), we created a list of 13 challenges related to mobile device usage. This list is not intended to be an exhaustive account of all possible barriers blind users may experience using a smartphone, but rather illustrates some of the most common ones:

\begin{enumerate}[label=C\arabic*]
\item{Performing a specific touchscreen gesture is hard} 
\label{c1}

\item{Unawareness of supported gestures in current app} \label{c2}

\item{Getting lost while using an app} \label{c3}

\item{Unable to scroll through lists} \label{c4}

\item{Unintentional selection of buttons} \label{c5}

\item{Smartphone is unresponsive} \label{c6}

\item{Screen reader is slow} \label{c7}

\item{Not knowing the meaning of a specific sound} \label{c8}

\item{Text input is slow} \label{c9}

\item{Changing written text is hard} \label{c10}

\item{Unable to login to apps and websites} \label{c11}

\item{Unable to find the desired option} \label{c12}

\item{Labels are inadequate or inexistent} \label{c13}
\end{enumerate}

We asked participants to rank-order these challenges accordingly to their perceived level of severity. We computed severity based on two dimensions [\cite{nielsen94}]: 1) frequency with which the problem occurs (is it common or rare?) and difficulty of the problem (is it easy or difficult to overcome the problem?).

Each challenge was listed on an RFID card (Figure \ref{figure:1}). We asked participants to rank-order the cards according to their relative perceived frequency or difficulty. Each dimension was ranked separately in a counterbalanced order. Each challenge was ranked from 1 (most frequent/difficult) to 13 (least frequent/difficult). Participants were free to adjust their rankings whenever they felt necessary during the study. They could also remove a card if they felt it did not represent a challenge for them.

\subsection{Apparatus}
Participants were each provided with a set of 13 RFID cards and an RFID reader (PROMAG PCR330). We developed a cross-platform web application that enabled participants to receive auditory feedback about each card. We used an HTML5-based Text-To-Speech library (responsivevoice.js) with a Portuguese female voice. Participants could touch a card with the reader and a laptop computer would read aloud the corresponding challenge.

\subsection{Procedure}
Participants were told that the overall purpose of the study was to understand the most important challenges they faced when using a smartphone. Following this, participants filled in a questionnaire about demographics and smartphone usage. 

We then explained that they would rank-order a set of challenges based on their personal experience using two criteria: frequency and difficulty. The rank-order for each of the two dimensions was conducted separately (one at a time) in a counterbalanced order. The deck of cards was shuffled after each ranking.

For ranking the frequency dimension, we directed participants to rank-order the 13 cards from the most frequent to the least frequent. We suggested participants to think about\textit{ "which of the following challenges occurs more frequently?"} On the other hand, for the difficulty dimension, we prompted:\textit{ "which of the following challenges is harder to overcome?"}. 

At the end of the study, we debriefed participants about their three top and bottom rankings in order to better understand their thoughts and decision rationale. In addition, participants rated the following statements on a 5-point Likert scale ranging from 1 (strongly disagree) to 5 (strongly agree): 1\textit{) "This activity was a fun way to think about smartphone challenges";} 2) \textit{"This activity was easy to do"}; 3. \textit{"This activity was fast to do"}; 4) \textit{"This activity was felt useless"}; and 5) \textit{"This activity helped me express my main concerns with smartphone usage"}

\subsection{Data Analysis}
To answer the first research question, which was to investigate the order of perceived importance of challenges, we first calculated the median scores of participants' ranking for each challenge in both dimensions (frequency and difficulty). We then ordered them with the lowest median score indicating the most frequent/difficult challenge. Finally, we conducted a k-means cluster analysis, with two clusters, to classify each challenge on a 2-dimensional diagram from low frequency/difficulty to high frequency/difficulty.

To answer the second research question, which was to examine whether there is an effect of experience in perceived importance of challenges, we applied nonparametric Mann-Whitney tests and calculated the correspondent r values. According to \cite{cohen88}, r values of .1, .3, and .5 represent small, medium, and large effect sizes, respectively. Although we collected information about the operating system of participants' device, we did not find any significant differences in rating between Android and iOS users.

To answer the third research question, which was to investigate blind users' views on the rank-ordering activity, we calculated the percentage of ratings on each score on a 5-point Likert scale. Nonparametric Mann-Whitney tests were applied to participants' ratings to explore whether there was a significant difference between novice and expert users.
\begin{table}[]
\centering
\resizebox{\textwidth}{!}{\begin{tabular}{llrrrrrr}
\hline
                                                      & \multicolumn{3}{l}{Overall} & \multicolumn{2}{l}{Experts} & \multicolumn{2}{l}{Novice} \\ \cline{2-8} 
Challenges - Frequency                                & Med (IQR)      & R    & NC  & Med (IQR)         & R       & Med (IQR)        & R       \\ \hline
C13: Labels are inadequate or inexistent              & 2 (3)          & 1    & 1   & 2 (1.5)           & 1       & 4 (3.25)         & 2       \\
C10: Changing written text is hard                    & 4 (5)          & 2    & 3   & 4.5 (3.75)        & 3       & 2 (5.25)         & 1       \\
C2: Unawareness of supported gestures in current app  & 5 (4.25)       & 3    & 1   & 8 (5.75)          & 10      & 5 (.5)           & 4       \\
C11: Unable to login to apps and websites*            & 5 (5.25)       & 4    & 3   & 3 (1.75)          & 2       & 8 (4.25)         & 10      \\
C3: Getting lost while using an app                   & 6 (4.25)       & 5    & 1   & 6 (3.25)          & 6       & 5.5 (4.5)        & 5       \\
C5: Unintentional selection of buttons                & 6 (5.5)        & 6    & 1   & 7 (5.75)          & 8       & 6 (5.5)          & 6       \\
C12: Unable to find the desired option                & 6.5 (3.25)     & 7    & 2   & 5 (2.5)           & 4       & 7 (3.75)         & 8       \\
C6: Smartphone is unresponsive                        & 6.5 (4.75)     & 8    & 0   & 6.5 (4.75)        & 7       & 6.5 (4.75)       & 7       \\
C1: Performing a specific touchscreen gesture is hard & 6.5 (8.25)     & 9    & 2   & 6 (4.5)           & 5       & 10 (8.25)        & 13      \\
C9: Text input is slow                                & 7 (10.25)      & 10   & 4   & 7 (7.25)          & 9       & 9 (7.5)          & 12      \\
C8: Not knowing the meaning of a specific sound       & 8.5 (8.25)     & 11   & 4   & 8.5 (9)           & 11      & 8.5 (6.75)       & 11      \\
C4: Unable to scroll through lists**                  & 8.5 (9.25)     & 12   & 6   & 10.5 (9.75)       & 13      & 4 (4.75)         & 3       \\
C7: Screen reader is slow                             & 8.5  (10.25)   & 13   & 4   & 10 (9.5)          & 12      & 7 (8)            & 9       \\ \hline \\
\end{tabular}}
\caption{Order of perceived frequency of challenges by expert and novice participants. Lower scores means higher frequency. Med = Median, IQR = Interquartile range, R = rank order based on median scores, NC = not a challenge. Challenges are ordered by overall ranking. Significant differences between groups are illustrated with * (p\textless.05) and ** (p\textgreater.1).}
\label{table:5} 
\end{table}
\subsection{Results}
In this section, we report results from the rank-ordering activity by analyzing the perceived frequency, perceived difficulty, and overall perceived importance of mobile challenges. Finally, we present participants' views on the rank-ordering activity itself.

\subsubsection{Perceived frequency of challenges}
Participants ordered each challenge by questioning themselves on \textit{"which of the following challenges occurs more frequently?"}. Median and interquartile ranges of expert and novice participants' frequency rankings are presented in Table \ref{table:5}. Overall, C13 (Labels are inadequate or inexistent) and C10 (Changing written text is hard) were perceived as occurring more frequently for both experts and novices. In fact they accounted for 50\% of the most frequent challenge (rank = 1);

Experts perceived C13 (\textit{Labels are inadequate or inexistent}), C11 (\textit{Unable to login to apps and websites}), and C10 (\textit{Changing written text is hard}) as the most frequent and C8 (\textit{Not knowing the meaning of a specific sound}), C7 (\textit{Screen reader is slow}), and C4 (\textit{Unable to scroll through lists}) as the least frequent. Similarly, novice participants perceived C13 and C10 as the most frequent but also mentioned C4 (\textit{Unable to scroll through lists}). C8 (\textit{Not knowing the meaning of a specific sound}), C9 (\textit{Text input is slow}), and C1 (\textit{Performing a specific touchscreen gesture is hard}) were perceived as the least frequent by novices.

Expert participants significantly perceived C11(\textit{Unable to login to apps and websites}) as occurring more frequently than novices (U=3, Z=2.589, p\textless.01, r=.72). According to \cite{cohen88} guideline, this was a large effect size. This result may be related with expert participants using more mobile/web applications and advanced features than novice users. Thus, considering the login process a frequent challenge. Additionally, we obtained a minor significant effect (U=4, Z=1.716, p=.086, r=.54) on C4 (\textit{Unable to scroll through lists}) with novice participants considering it a challenge more frequently (rank 3 vs rank 13).

During the rank-ordering procedure, participants were allowed to remove cards if they did not considered them challenges. Expert participants removed 18 cards while novices removed 14 cards (no significant differences, Z=.674, p=.5). As illustrated in Table \ref{table:5}, C4 was not considered a challenge by 6 out of 16 participants, followed by C7, C8, and C9 with 4 participants classifying them as not challenges. Interestingly, these were also the least frequent challenges.

\subsubsection{Perceived difficulty of challenges}
In this stage of the user study, participants ordered challenges while questioning themselves \textit{"which of the following challenges is harder to overcome?"}. Table \ref{table:6} shows the median and interquartile ranges of expert and novice participants' difficulty rankings. Similarly to frequency, C13 (\textit{Labels are inadequate or inexistent}) and C10 (\textit{Changing written text is hard}) were perceived as overall more difficult accounting for 56\% of the most frequent challenge (rank = 1);

\begin{table}[]
\centering
\resizebox{\textwidth}{!}{\begin{tabular}{llrrrrrr}
\hline
                                                      & \multicolumn{3}{l}{Overall} & \multicolumn{2}{l}{Experts} & \multicolumn{2}{l}{Novice} \\ \cline{2-8} 
Challenges - Frequency                                & Med (IQR)  & R  & NC      & Med (IQR)  & R      & Med (IQR)  & R  \\ \hline
C13: Labels are inadequate or inexistent              & 2 (3.75)   & 1  & 1       & 1.5 (3)    & 1      & 3 (6.25)   & 3  \\
C10: Changing written text is hard*                   & 3 (3.5)    & 2  & 3       & 8.5 (9)    & 10     & 3 (2.25)   & 1  \\
C12: Unable to find the desired option                & 3 (5.5)    & 3  & 2       & 3 (1)      & 2      & 9 (7.25)   & 12 \\
C11: Unable to login to apps and websites             & 4 (4.25)   & 4  & 3       & 5.5 (4.5)  & 5      & 3 (4)      & 2  \\
C3: Getting lost while using an app                   & 5 (2.25)   & 5  & 1       & 5 (2)      & 3      & 5.5 (3.25) & 4  \\
C6: Smartphone is unresponsive                        & 6 (6.25)   & 6  & 0       & 5 (7)      & 4      & 7 (4)      & 9  \\
C9: Text input is slow                                & 6.5 (5.75) & 7  & 4       & 9 (8.25)   & 11     & 6 (4)      & 6  \\
C2: Unawareness of supported gestures in current app  & 7 (5)      & 8  & 1       & 7 (4.25)   & 8      & 6.5 (5.25) & 8  \\
C8: Not knowing the meaning of a specific sound       & 7 (6.75)   & 9  & 4       & 6.5 (6.25) & 7      & 8 (5.75)   & 10 \\
C1: Performing a specific touchscreen gesture is hard & 7 (7.25)   & 10 & 2       & 6 (6)      & 6      & 9 (5.75)   & 11 \\
C4: Unable to scroll through lists                    & 7 (8.25)   & 11 & 6       & 9 (9)      & 12     & 6 (2.75)   & 5  \\
C5: Unintentional selection of buttons                & 8 (4.25)   & 12 & 1       & 8 (2)      & 9      & 6 (5)      & 7  \\
C7: Screen reader is slow                             & 10 (10.75) & 13 & 4       & 10 (9.5)   & 13     & 10.5 (10)  & 13 \\ \hline \\
\end{tabular}}
\caption{Order of perceived difficulty of challenges by expert and novice participants. Lower scores means higher frequency. Med = Median, IQR = Interquartile range, R = rank order based on median scores, NC = not a challenge. Challenges are ordered by overall ranking. Significant differences between groups are illustrated with * (p\textless.05).}
\label{table:6}
\end{table}

Experts perceived C13 (\textit{Labels are inadequate or inexistent}), C12 (\textit{Unable to find the desired option}), and C3 (\textit{Getting lost while using an app}) as the most difficult challenges to overcome and C9 (\textit{Text input is slow}), C4 (\textit{Unable to scroll through lists}), and C7 (\textit{Screen reader is slow}) as the least difficult. Novice participants perceived C10 (\textit{Changing written text is hard}), C11 (\textit{Unable to login to apps and websites}), and C13 as the most difficult and C1 (\textit{Performing a specific touchscreen gesture is hard}), C12 (\textit{Unable to find the desired option}), and C7 (\textit{Screen reader is slow}) were perceived as the least difficult.

Novice participants significantly perceived C10 (\textit{Changing written text}) as more difficulty to overcome than experts (U=6, Z=2.182, p\textless.05, r=0.6, large effect). The result suggests that expert users develop coping strategies over time to edit written text. In fact, in the debriefing session, some participants mentioned using external devices to perform text-entry tasks, such as Braille notetakers. These devices ease editing operations such as cursor movement, text select, and clipboard operations; however, they require pairing Bluetooth devices, and knowing how to operate the external keyboard, which may pose new barriers to novice users. 

\subsubsection{Perceived importance of challenges}
In this section, we examine the perceived importance of mobile challenges taking into account both median frequency and median difficulty. We conducted a k-means clustering analysis (with two clusters) to group similar challenges on each dimension, which resulted in four quadrants: high frequency and high difficulty, high frequency and low difficulty, low frequency and high difficulty, and low frequency and low difficulty. Notice that the most important challenges will be in the high frequency and high difficulty quadrant; these are challenges that occur frequently and are hard to overcome. On the other hand, challenges that are perceived as the least important will be in the low frequency and low difficulty quadrant.

\begin{figure}
\centering
\begin{minipage}{.5\textwidth}
  \centering
  \includegraphics[width=\linewidth]{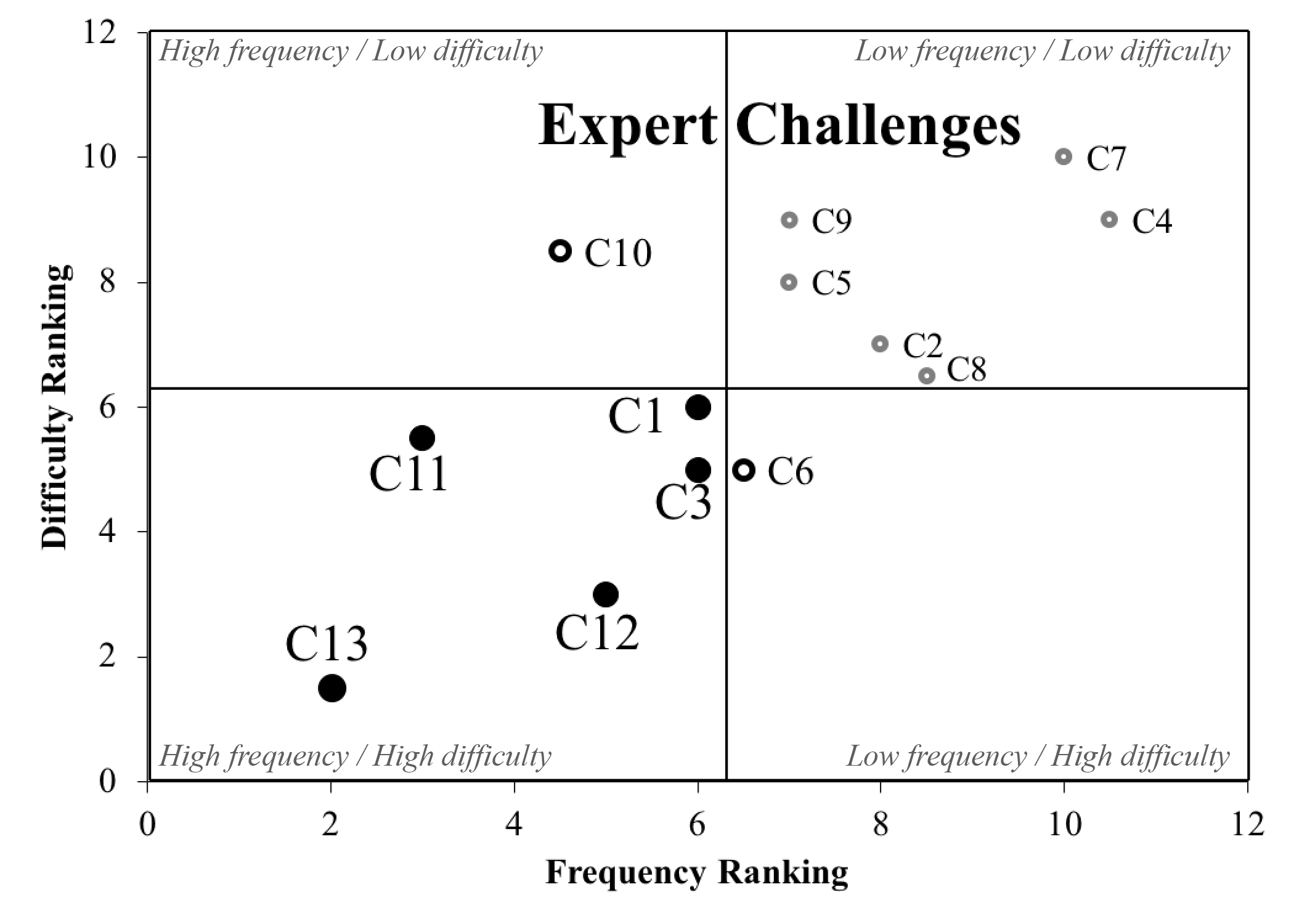}
  \caption{Quadrant analysis for expert participants}
  \label{fig:2}
\end{minipage}%
\begin{minipage}{.5\textwidth}
  \centering
  \includegraphics[width=\linewidth]{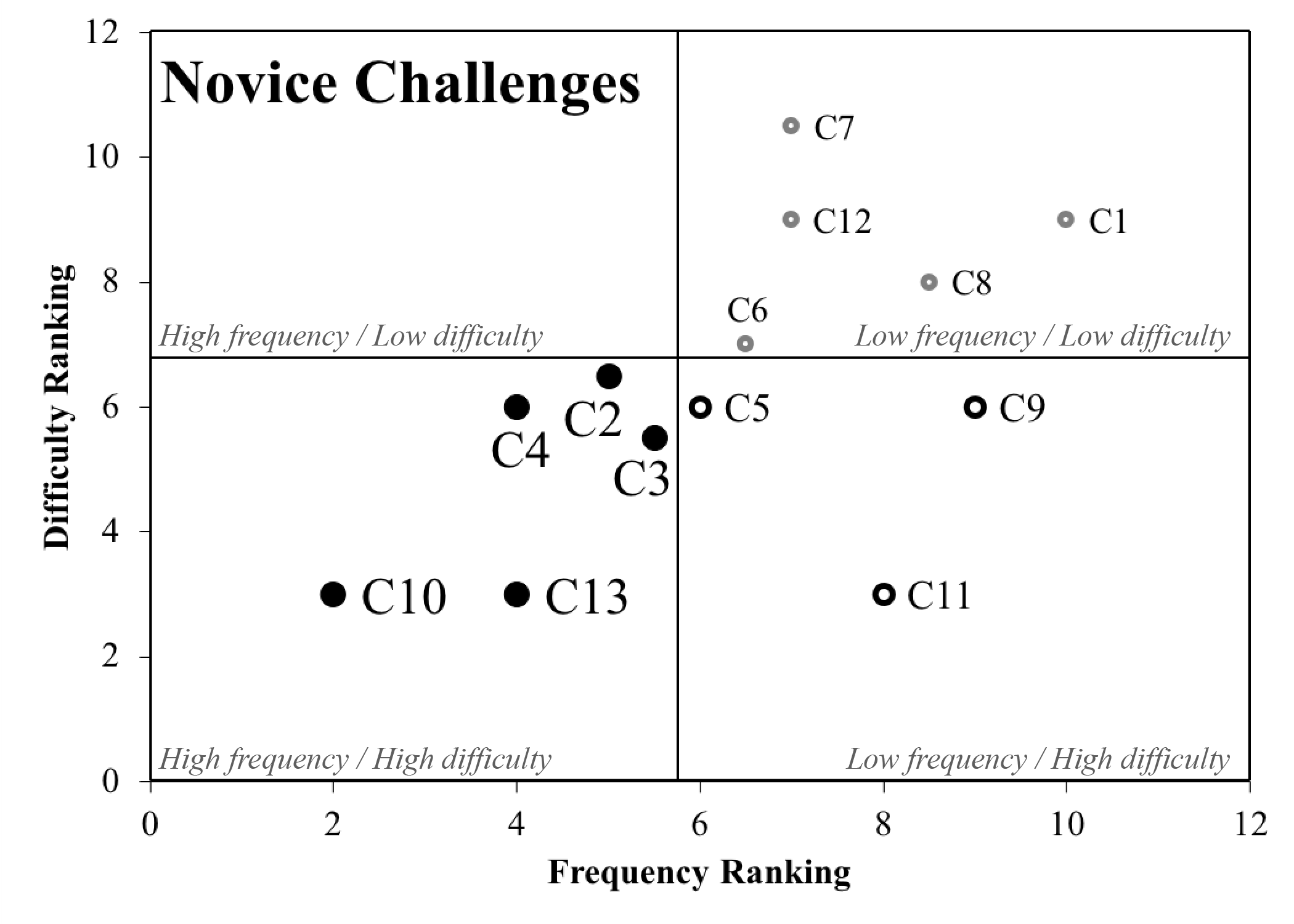}
  \caption{Quadrant analysis for novice participants.}
  \label{fig:3}
\end{minipage}
\end{figure}

Overall, the most important challenges were C11 (\textit{Unable to login to apps and websites}), C10 (\textit{Changing written text is hard}), and C13 (\textit{Labels are inadequate or inexistent}). Figures \ref{fig:2} and \ref{fig:3} show the quadrant results for expert and novice participants, respectively.

Expert participants perceived C1 (\textit{Performing a specific touchscreen gesture is hard}), C3 (\textit{Getting lost while using an app}), C11 (\textit{Unable to login to apps and websites}), C12 (\textit{Unable to find the desired option}), and C13 (\textit{Labels are inadequate or inexistent}) as the most important. C10 (\textit{Changing written text is hard}) and C6 (\textit{Smartphone is unresponsive}) are still relevant challenges either because they occur frequently or are hard to solve.

Regarding novice participants, the most relevant challenges are C2 (\textit{Unawareness of supported gestures in current app}), C3 (\textit{Getting lost while using an app}), C4 (\textit{Unable to scroll through lists}), C10 (\textit{Changing written text is hard}), and C13 (\textit{Labels are inadequate or inexistent}). C5 (\textit{Unintentional selection of buttons}), C11 (\textit{Unable to login to apps and websites}), and C9 (\textit{Text input is slow}) were classified as occurring with less frequency but still hard to overcome.

\subsubsection{Views on ranking activity}
To the best of our knowledge, we were the first to explore a rank-ordering activity using a non-visual approach with RFID tags. Thus, we were interested in assessing participant's views on the activity itself.

\begin{table}[]
\centering
\resizebox{\textwidth}{!}{\begin{tabular}{lccccc}
\hline
Item / Rating                                                      & 1       & 2       & 3       & 4       & 5       \\ \hline
This activity was easy to do                                       & 0       & 0       & 31.30\% & 18.80\% & 50\%    \\ \hline
This activity was fast to do                                       & 0       & 0       & 12.50\% & 37.50\% & 50\%    \\ \hline
This activity was a fun way to address mobile phone challenges     & 0       & 0       & 12.50\% & 25\%    & 62.50\% \\ \hline
This activity was useless                                          & 81.30\% & 18.70\% & 0       & 0       & 0       \\ \hline
This activity help me express my biggest problems with smartphones & 0       & 0       & 12.50\% & 25\%    & 62.50\%  \\ \hline  \\ 
\end{tabular}}
\caption{Participants' ratings to each questionnaire item.}
\label{table:7}
\end{table}

The majority of participants (87.5\%) were positive that the activity was fun and helped them express the challenges they experience with smartphones (Table \ref{table:7}). In addition, 68.8\% indicated that the activity was easy and   87.5\% fast to do. Nearly one third of participants had a neutral view on how easy it was to perform the activity. This result may be related with the mental effort of remembering the position of all cards. Although participants were able to read the cards freely, they often tried to remember the ordering to correctly place a new card. Thus, future researchers should keep the number of cards to a manageable number. All participants participants disagreed that the activity was useless. Finally, both expert and novice participants had positive views on the rank-ordering activity, and we found no significant differences between user groups.

\subsection{Discussion}
Overall, these results suggest that physical card sorting is a feasible, useful, and enjoyable approach that provide blind users the opportunity to express their opinions and facilitate communication. 

\subsubsection{Novice users struggle with fundamentals to operate the device}
Novice participants gave the highest importance to\textit{ Unawareness of supported gestures in current app, Unable to scroll through lists,} and \textit{Changing written text is hard}. Similar to expert participants, they also prioritised \textit{Getting lost while using an app} and \textit{Labels are inadequate or inexistent}. Novice users emphasised the importance of basic operations needed to fully control a smartphone. Although most of the aforementioned challenges are addressed in Android and iOS introductory tutorials, results suggest that they may not be an effective learning tool. There has been previous research that offered insight into the challenges of non-visual text editing \cite{azenkot13,Trindade18}. Azenkot et al. \cite{azenkot13}  investigated the challenges of using speech input on mobile devices and found that while participants were mostly satisfied with the input modality, editing entered text was frustrating. Participants spent 80\% of their time correcting errors. \cite{Trindade18} showed that performing selection and clipboard operations are the most challenging and time consuming as users need to navigate through several menus. Although the authors propose a new editing technique, it focuses on Braille input and requires additional hardware.

There have been several research projects aiming to support non-visual navigation through lists of items \cite{Kane09, McGookin08,ahmed10}, which have been adopted by mainstream screen readers. However, it has been assumed that these techniques are easy to learn and use. Moreover, most research focus on improving non-visual gestural input rather than supporting the discoverability of such gestures.

\subsubsection{Expert users struggle with app-specific functionality}
In the present study, expert participants gave the highest importance to \textit{Performing a specific touchscreen gesture is hard, Unable to login to apps and websites, and Unable to find the desired option}. Challenges experienced by expert users are frequently related with functionality rather than basic operations. For example, the login process is perceived as challenging. Although previous research has highlighted the unique privacy and security needs of blind people \cite{ahmed2015} and several eyes-free authentication methods have been proposed \cite{AzenkotRector12,sun14}, users still struggle with traditional login methods either because they are inaccessible or hard to use.

Results show that gestural input is sometimes hard, particularly for specific gestures. Previous research has investigated the most appropriate gestures for blind people \cite{Kane09} and shown that Android's L-shaped gesture or iOS rotor gesture can be troublesome \cite{Rodrigues15}. 

\subsubsection{Everyone gets lost}
Both novice and expert participants ranked \textit{Getting lost while using an app} high in terms of frequency and difficulty. From the debriefing session, participants mentioned that learning to use a new app usually takes time and requires persistency. It is common to frequently get lost while trying find a specific functionality or trying to do something new. App updates are also an issue, forcing people to rebuild their mental model about the apps' structure and behaviours. The learning process is usually done through serendipitous exploration and trial-and-error. When users get lost using an app they will often rely on completely closing and starting the app again; or a hard reset of the device to resurface at a known state within the app.

\newcommand*{\MyIndent}{\hspace*{0.8cm}}%
\newcommand*{\MyIndentt}{\hspace*{0.4cm}}%
\newcommand*{\MyIndenttt}{\hspace*{0.6cm}}%

\begin{table}[]
\resizebox{\textwidth}{!}{%
\centering
\begin{tabular}{@{}ll@{}}
\toprule
Category & Challenge  \\ \midrule
1. Gestures           & 1.1 Perform a specific touchscreen gesture\\
       \MyIndent \&    &  \MyIndent 1.1.1 Learn a new gesture\\
     \MyIndentt Navigation     &  \MyIndent 1.1.2 Perform a gesture consistently correct\\
           & 1.2 Perform unintentional gestures\\
           & 1.3 Scroll through lists\\
          \midrule
2. Awareness           & 2.1 Find the desired option\\
                  & 2.2 Be aware of the available options\\
                            & 2.3 Be aware of supported gestures\\

                    \midrule

3. Mental Model          
		& 3.1 Get lost when using an app\\
          & 3.2. Unexpected navigation triggered without performing an action\\
          & 3.3 Return to the first screen of an app\\
          & 3.4 Understand content structure\\
          & 3.5 Understand 	features and their effects\\
          & 3.6 Adapt to an update\\
          & 3.7 Overwhelmed by features\\
          \midrule
4. Feedback          & 4.1 Smartphone is unresponsive\\
          & 4.2 Screen reader is slow\\
          & 4.3 Not knowing the meaning of a specific sound\\
          & 4.4 Screen reader is too verbose\\
          & 4.5 Overlapped feedback\\
          \midrule
5. Text          & 5.1 Text input is slow\\
          		& 5.2 Editing text\\
                          \midrule
6. Security          & 6.1 Login in apps and websites\\
	          & 6.2 Unlocking the device\\
                        \midrule
7. Accessibility          & 7.1 Labels are inadequate or inexistent\\
          & 7.2 Unreachable elements\\
                    \midrule
8. Hardware          & 8.1 Configuring external devices to connect to the smartphone\\
          & 8.2 Using capacitive buttons\\
              \midrule
9. Sharing           & 9.1 Consistently reproduce an issue\\
   \MyIndent \&           & 9.2 Others do not know how to use my device\\
\MyIndentt Assistance          & 9.3 Be aware of what others are doing when handling my device\\
          & 9.4 Restrict others access to one's personal device\\
          & 9.5 Find knowledgeable assistance\\
              \midrule
10. Knowledge           & 10.1 Know which device to purchase\\
  \MyIndent Void           & 10.2 Find fully accessible apps\\

 \midrule 
\end{tabular}}
\caption{Smartphone challenge corpus.}
\label{table:8}
\end{table}

\section{Outlook on Smartphone Accessibility}
 Smartphone accessibility is currently in a dichotomy of states. We can consider smartphones accessible because accessibility services (e.g. screen readers) allow users to reach every piece of content and many applications are fully accessible in the sense that every element is identified. Moreover, we have people that have been able to adopt the device and become proficient users able to tackle almost any issue thrown at them.  However, for many, the current state of affairs is not enough. People struggle to adopt the device and continue to face challenges at every level of expertise.  Moreover, two blind people using the same interface can have a vastly different experience depending on the interaction behaviors, ability, and desire to go through a trial and error process.  It is now evident by these studies the lack of support people  have when starting to use the device and how much they have to rely on others to overcome the frailties of the process. For many without an established network of support this leads to either never transitioning to a smartphone, or forever remaining a user of basic phone tasks. There are currently no support mechanism that guide and help the user evolve its smartphone usage. 

Once a user goes through the adoption process the challenges do not cease, they simply morph. Newcomers struggle to establish mental models and are overwhelmed by the variety of behaviors, interfaces, and feedback mechanisms. For experienced users,  new apps and updates  create the need to adapt, rediscover and commit entire new interfaces to memory in order to efficiently use their smartphone.  None of this would be a problem if all it took to adapt was a quick glance. However, the equivalent for blind people is a long process of hearing everything in the screen and possibly interacting with a few elements to  assess the changes and their impact.  Another consequence of the variety of interfaces layouts and behaviors is the lack of awareness of available options that plagued users of all expertise levels. 

The number and variety of challenges users have to overcome, lead them to different coping methods. Some neglect the device and only rely on it for simple tasks. Others are still carefully handling old feature phones in an effort to prevent the frustration and self-doubt on ones ability that comes with struggling to operate a mobile device, especially one that for all intents  and purposes can be accessible. The social pressure of knowing people around them were able to adapt can be damaging when they first start and struggle to learn. The most common coping method is relying on friends and family whenever a challenge appears, which is far from ideal. For the few that are tech savvy, online help from community forums and other platforms (e.g. YouTube) resolves most of their issues, with the remaining requiring persistence - having the desire and ability to learn through trial and error. 

Despite all efforts in mobile accessibility, smartphones are still not accessible to all. People face a variety of challenges that prevent some from taking  advantage of their device features and make others dependent on their support network. 
 
 In this work, we identified and verified a large set of challenges experienced by blind people (Table \ref{table:8}). We reported on some insights into the current coping mechanisms people rely on. Finally, we prioritized a sub-set of some of the most common ones. Below we discuss the the impact of relying on a multi-method approach. 

\subsection{Multiple-Methods Approach}
We presented three user studies where we relied on a variety of data sources and collection methods to have a holistic view of the challenges faced and coping mechanisms employed by blind people when interacting with smartphones. In \cite{Rodrigues15} we relied on a mixed-method approach that allowed us analyze, in great detail, the evolution that our five participants went through. In this paper first study, we reached a wider participant pool and realized that locally, the majority of users were newcomers or novice. Moreover, we understood that the learning process never stopped and some remained at a novice level despite years of experience due to the lack of support mechanisms.  We also identified a select few that  take advantage of all the device has to offer. The second study led us to question the challenges faced by the particular user group of expert tech savvy users. We became interested in understanding how these users addressed their issues.  In this paper second study, we focused on understanding the issues depicted in the top online community forums giving us yet another view of the challenges and coping mechanisms that can be effective for a segment of the user group.  Finally, our last study enabled us to assess the severity of each of the challenges faced as perceived by users of different expertise levels.  

Had we limited our research to a single laboratory or in-the-wild study, we would only have a fragmented view of the problem capturing only snapshots of the challenges faced. Without multiple methods we would not have captured the full spectrum of the smartphone challenges from newcomers to experts. Understanding how challenges morph and how different users tackle the same issues differently. When understanding users behaviors and challenges with technology, every study raises questions that should be pursued. Through a multi-method approach we were able to delve wide and deep into the challenges blind people have when interacting with smartphones. The approach allowed us to understand the adoption process, the variety of expertises, the challenges, the coping mechanisms, the different impact of different stakeholders, and users' knowledge progression.  

We urge researchers that are exploring challenges in technology adoption and usage  to take a multi-method approach providing a holistic view that encapsulates all types of users' challenges, coping mechanisms and identifying the different impact factors and thus, possible areas for improvement.  

\section{Directions for Future Research}
In this work we identified a set of open challenges. We assessed their severity as perceived by participants based on frequency and difficulty. Several of the identified issues have been previously reported and heavily researched (i.e. inadequate labels \cite{Takagi08}, inputing text \cite{Rodrigues16,Nicolau17,azenkot13} learning and performing gestures \cite{Oh13}). However, many others have been largely overlooked or under explored (i.e. getting lost while using an app, unawareness of supported gestures, unable to find desired option). Bellow we discuss how identified challenges categories still have open questions and posit areas of attention for future research in mobile accessibility:

\subsection{Learning and Discovering} 
Although there has been some work on how to  learn \cite{Oh13} and perform gestures on touchscreens \cite{Kane11, McGookin08}, it appears the challenges are not yet surpassed. Gestures are not discoverable nor easy to learn based on some of the given descriptions and users have no fallback mechanism to rely on, except for their support network. Perhaps it is time to think about how  gesture discovery and practice can be embedded into everyday interactions, how can we track performance, hint at corrections,  adapt recognizers or even develop entertainment apps whose sole goal is gesture practice and discovery. For instances, one could develop a game to for users to learn and train screen reader gestures, similar to what Microsoft did the mouse with the release MineSweeper for Windows 3.1 \cite{kapp2012gamification}.
Awareness of the available options was one of the most reported challenges. Past work in multiple audio sources \cite{Guerreiro15, Guerreiro14} has explored how to augment awareness in text input and when passively consuming news. There is an opportunity to leverage users ability to segregate multiple audio sources and explore novel audio navigation techniques that can be applied to other tasks.  

\subsection{Adapting Mental Models} 
The fast paced, iterative nature of the devices' OS, and apps bring about additional complexities. Interfaces can be radically reinvented with each new OS update or subsequent version of an app. Meaning that users must also continuously rediscover and adapt their mental models of interaction to maintain their current level of expertise and ability. However, there has been limited work exploring the design space of assisting blind people with these challenges on mobile devices. For example, how can we inform users of app interaction changes that are relevant for their usual workflows. Is it possible to enable the voice assistant to also be an app assistant? There is an opportunity to explore novel methods that guide, inform, and promote smartphone usage on a continuous basis beyond first steps.

\subsection{Everyday Accessibility Issues} 
Despite the vast amount of work that has targeted basic accessibility challenges they are still prevalent and relevant. Previous research has  proposed a set testable guidelines and evaluated  the most popular 25 apps from Play Store and found a widespread of accessibility issues  \cite{Ballantyne2018}.  There have been recent in standardization efforts with the W3C released a of the  Web Content Accessibility Guidelines 2.1 that now take into further account the specificity's of mobile devices. Some accessibility efforts have also been industry 
led, particularly for native applications (e.g. Google Material Guidelines\cite{google_material} , iOS Guidelines \cite{appleguidelines}, BBC Mobile Accessibility Guidelines  \cite{bbc}).  However, the smartphone ecosystem is highly fragmented with no guidelines currently imposed on developers.  Recent work by  \cite{Ross:2017:EFL:3132525.3132547} propose an accessibility framework based on epidemiology, seeking to explore the accessibility issues of mobile apps at a population level within the context of the entire mobile (and software development) ecosystem over time. There is work to be done in how technology can intervene at development time to ensure compliance with accessibility requirements. For instances, common building blocks for interfaces should be accessible as Apple has done for some of their pervasive controls and widgets. 

Although there has been significant research conducted in text-entry current mechanisms are still difficult to use effectively. Users often discuss alternatives to the current status quo of virtual QWERTY keyboards and rely on braille base input or external keyboards for faster input rates. Text-editing and clipboard operations are still out of reach for most and only recently have we seen efforts in improving current methods \cite{Trindade18}. 

\subsection{Forms no Standards one Account} 
The recent trend to use one account to register and login (i.e. log in with Google/Facebook) has the potential to minimize the issues faced with registrations and logins. However, there are consequences to relying on a single entity and to allow apps to have access to additional user information through these accounts. Unlike sighted users who rely on this forms of registration for time savings, for blind people they might be the only way to be able to successfully complete a registration within reasonable time. Applications and websites are free to create their forms as they wish, there is no standardization of how forms are presented nor currently a way to adapt them. Further work in accessibility proxies \cite {zhanginteraction} may be one way to address these issues. 

\subsection{Forced Interfaces} 
The decline of physical buttons is still a concern for some who wish to turn back the clock. For many people, adapting to touchscreen devices is a need that was forced upon by the shift in the phone market. Although people recognize the benefits of touchscreens, for most, they are still not efficient or as easy to handle when compared with physical keys. Providing assistance can be beneficial for people who struggle to interact with touchscreens, but it won’t replace the need for better interaction methods. There has been some work in augmenting current devices with cases for multiple purposes, from having a t9 \cite{Zhang:2018:IPT:3234695.3236349}, to editing text \cite{Trindade18}. However, blind people do not require to have the smartphone out to receive any feedback. Is a touchscreen really the best input method we can design for blind people to interact with technology on the go? We have this artificial restriction imposed solely because sighted people require a screen for feedback.  For desktops, televisions, entertainment systems and other technology, it seems that often touchscreens are the least accessible option. What can we design if we think about taking a 'controller' out of the pocket instead of the smartphone? We believe there is an opportunity to design novel interaction controllers that enable users to reap the benefits of smartphones, while not being restrained by the touchscreen.  

\subsection{Ubiquitous Accessibility Information}
During our studies people expressed the lack of available information regarding the accessibility of different devices and applications. Discovering and exploring new apps is always a daunting task, where users are often faced with completely inaccessible applications (e.g. no labels). Similarly, updates and changes are frequent. When they happen, users have to re-adapt to already known interfaces. Past research has already started to investigate how the choice of technology, or the uptake of new ones, can impact the overall accessibility in the context of web pages \cite{richards}. We believe there is an entire field of underexplored research going beyond understanding how app accessibility evolves, and how it affects each and all individuals. We need to start acting on this knowledge and develop solutions that are designed to address the volatility of today, providing the relevant information to the users at the different stages. 

Users struggle to navigate the device and app market with no knowledge available of their accessibility features or compliance.  In both instances users have to rely on third parties (e.g. friends, forums) to know about the device or apps. For mobile applications users can leverage ratings, app stores descriptions and comments to try and pre-assess their accessibility. A dedicated accessibility rating and other metrics could allow users to make informed decisions. Automatic evaluations and new metrics for accessibility may be needed to accommodate the dynamic nature of most mobile apps that frequently cause awareness and mental model challenges. 

\subsection{Enable Sharing and Peer Support} 
We found that not only is the adoption process a long and arduous task, but there are limited tools to guide and support users through this process. The status quo relies on users persistence and ability to seek aid from others to overcome the challenges presented.  Regardless of how well a technology is designed there will always be times when additional peer support is needed. Through our workshops and interviews we heard tales of family and loved ones tirelessly answering  phone calls and requests for assistance and guidance to everyday tasks - with individuals asked the same question time and time again. In recent years we've seen innovative solutions that leverage crowdworkers to assist disabled people overcome accessibility challenges \cite{Bigham10}. While these works have demonstrated that small micro-tasks and contained questions can be easily handled by crowdworkers, there is still a need for longer, more engaged and curated support such as one-to-one walk-throughs and tutorials. More experienced users were able to leverage existing technologies i.e asking questions in on-line community forums. However, many others were unable, or found the information inaccessible or in-compatible with their device configuration (Hardware, OS and App versions). Through our own first attempts to provide in-context question \& answer support \cite{Rodrigues17}, we identified significant limitations with the existing communication methods for such support. We believe there is an opportunity for new forms of assistance that don't require the user to take additional complex steps; that are tailored to their individual needs; and can be readily available when it is next needed. To achieve such assistance we must also empower the support networks with the understanding and tools that will allow them to communicate this assistance in an accessible manner. 

Mobile screen readers are not intuitive for sighted people, nor they were designed to be. However, as we have reported, sharing and assistance are common occurrences that are currently hampered. There is work to be done in enabling sighted users to interact with screen readers as well as in creating mechanisms by which blind people are aware and in control of such interactions. Users should not have to trust sighted peers to solve issues or when sharing something on the device.    

\subsection{Limitations}
In this research we have taken a multiple-methods approach, whereby the subsequent user studies build on and overlap with the prior user engagements in an attempt to verify the outcomes and our conclusions drawn. Nevertheless, we would like to highlight that results show the relative importance of challenges. Lower ratings do not mean that challenges are not important to address; they are just perceived as relatively less important. 

\section{Conclusion}
Through this work we explored the everyday challenges encountered by blind people with smartphone devices. Our multiple-method approach, combining data from real-world usage logs; interviews; ethnographic observations and workshops with interface evaluations and non-visual card ranking, has allowed us to not only uncover the realities of non-visual interaction smartphone challenges, but to prioritise those issues based on their perceived frequency of occurrence and impact on difficulty. In addition to the detailed description of our methodological approach, we contribute a corpus of smartphone challenges as experienced by blind people, validated through different lenses; and an in-depth discussion of open directions for future research.

\section{Acknowledgments}
We thank Funda\c{c}\~ao Raquel and Martin Sain in Lisbon (Portugal) and all participants. This work was partially supported by  Funda\c{c}\~ao  para a Ci\^encia e Tecnologia (FCT) through scholarship SFRH/BD/103935/2014, INESC-ID research unit  UID/CEC/50021/2019, and LASIGE research unit UID/CEC/00408/2019 and by DERC EP/M023001/1 (Digital Economy Research Centre).


\bibliographystyle{unsrt}  


\end{document}